\documentclass{jps-cp}
\usepackage{txfonts} 
\usepackage{footmisc}

\title{$\Omega_c$ excited states: \\ a molecular approach with heavy-quark spin symmetry}

\author{Tolos \textsc{Laura}$^{1,2,3,4}$, Pavao  \textsc{Rafael}$^{5}$  and Nieves \textsc{Juan}$^{5}$}

\inst{$^{1}$ Institut f\"ur Theoretische Physik, University of Frankfurt, Max-von-Laue-Str. 1, 60438 Frankfurt am Main, Germany \\
$^{2}$ Frankfurt Institute for Advanced Studies, University of Frankfurt, Ruth-Moufang-Str. 1, 60438 Frankfurt am Main, Germany \\
$^{3}$ Institute of Space Sciences (ICE, CSIC), Campus UAB, Carrer de Can Magrans, 08193, Barcelona, Spain\\
$^{4}$ Institut d'Estudis Espacials de Catalunya (IEEC), 08034 Barcelona, Spain\\
$^{5}$  Instituto~de~F\'{\i}sica~Corpuscular~(centro~mixto~CSIC-UV), Institutos~de~Investigaci\'on~de~Paterna, Aptdo.~22085,~46071,~Valencia,Spain}

\email{tolos@th.physik.uni-frankfurt.de}

\recdate{December 14, 2018}

\abst{The LHCb Collaboration has recently discovered five excited $\Omega_c$ states with masses between 3 and 3.1 GeV, four of them corroborated by the Belle Collaboration. We analyse the dynamical generation of these states within a molecular baryon-meson model that is consistent with both chiral and heavy-quark spin symmetries. Earlier predictions within this model found five $\Omega_c$ states with masses below 3 GeV. Thus, in order to study the possible identification of any of these states with the experimental ones in the correct energy region, we explore two different regularization schemes, that is,  a modified regularization subtraction method and a cutoff regularization scheme. We find that at least three of the dynamically generated states can be identified with the experimental ones and have spin-parity $J=1/2^-$ or $J=3/2^-$}

\kword{$\Omega_c$ excited states, heavy-quark spin symmetry, molecular model}

\begin{document}
\maketitle

\section{Introduction}
\label{intro}

Five  excited $\Omega_c$ states with masses between 3 and 3.1 GeV have been recently discovered by the LHCb Collaboration \cite{Aaij:2017nav} in the $\Xi_c^+ K^-$
decay in $pp$ collisions, being four of them corroborated by the Belle Collaboration \cite{Yelton:2017qxg}.   Predictions on these states have been consequently revisited within quark models, QCD sum-rule schemes, quark-soliton models,
lattice QCD and molecular models, so as to understand whether these states can be explained within the conventional quark model picture and/or these states are of molecular type.

Earlier predictions of excited $\Omega_c$ states within molecular models \cite{JimenezTejero:2009vq,Hofmann:2005sw,GarciaRecio:2008dp, Romanets:2012hm} have been indeed reanalyzed in view of the new discoveries. Whereas in Ref.~\cite{Montana:2017kjw} two $\Omega_c$ resonant states  at 3050 MeV and 3090 MeV with $J^P=1/2^-$ were obtained,  being identified with  two of the experimental states, the extended local hidden gauge approach of Ref.~\cite{Debastiani:2017ewu} predicted  two  $J^P=1/2^-$ $\Omega_c$ states  and one $J^P=3/2^-$ $\Omega_c^*$ , the first two in good agreement with \cite{Montana:2017kjw}.

However, some of these previous molecular models do not respect heavy-quark spin symmetry (HQSS), which is a proper QCD symmetry in the limit of large quark masses beyond the typical confinement scale. Therefore,  a scheme that explicitly includes HQSS has been developed over the past years \cite{GarciaRecio:2008dp,Gamermann:2010zz,Romanets:2012hm,GarciaRecio:2012db,Garcia-Recio:2013gaa,Tolos:2013gta,Garcia-Recio:2015jsa}. This model is based on a ${\rm SU(6)}_{\rm lsf} \times {\rm HQSS}$ extension of the Weinberg-Tomozawa (WT) interaction, with "lsf'' referring to light quark-spin-flavor symmetry. Indeed,  Refs.~\cite{GarciaRecio:2008dp, Romanets:2012hm,GarciaRecio:2012db,Garcia-Recio:2015jsa} are the first  baryon-meson molecular analyses, fully consistent with HQSS, that reproduce the odd-parity $\Lambda_c(2595)$ [$J=1/2$] and $\Lambda_c(2625)$ [$J=3/2$] resonances \cite{GarciaRecio:2008dp, Romanets:2012hm,Garcia-Recio:2015jsa} as well as the $\Lambda_b(5912)$ [$J=1/2$] and $\Lambda_b(5920)$ [$J=3/2$] narrow resonances \cite{GarciaRecio:2012db,Garcia-Recio:2015jsa},  these last two found by LHCb \cite{Aaij:2012da}.

In Ref.~\cite{Romanets:2012hm} five $\Omega_c$ states were generated dynamically, three $J=1/2$ and two $J=3/2$ bound states,  that were stemming from the most attractive ${\rm  SU(6)}_{\rm lsf}\times$ HQSS representations and with masses below than those predicted by the LHCb. However, the prediction of the masses  strongly depends 
on the adopted regularization scheme (RS). In this work we revisit the RS used in \cite{Romanets:2012hm}, so as to reanalize the five dynamically generated $\Omega_c^{(*)}$ states. We find that at least three states can be identified experimentally by implementing a modified RS \cite{Nieves:2017jjx}.

\section{Molecular model with heavy-quark spin symmetry}
\label{unitarized}

We reanalize the work of Ref.~\cite{Romanets:2012hm} by studying the dynamical generation of the five $\Omega_c^{(*)}$ excited states located in the charm $C=1$, strangeness $S=-2$ and isospin $I=0$ sector. The $s$-wave baryon-meson potential $V^J$ (for a total angular momentum $J$) results from the  ${\rm SU(6)}_{\rm lsf}$ $\times$ HQSS WT interaction that involves pseudoscalar and vector mesons together with the low-lying $1/2^+$ and $3/2^+$ baryons. Solving  the Bethe-Salpeter equation in the on-shell approximation, we obtain the scattering amplitude  ($T$-matrix) 
\begin{equation}
\label{eq:LS}
T^J(s)=\frac{1}{1-V^J(s) G^J(s)} V^J(s),
\end{equation}
with the diagonal $G^J(s)$ matrix containing the baryon-meson loop functions. The  loop function is
logarithmically ultraviolet (UV) divergent, and it has to be
regularized. That is, the loop function for each channel $i$ is given by
\begin{equation}
G_i(s)=\overline{G}_i(s)+G_i(s_{i+}) ,
\label{eq:div}
\end{equation}
with a finite part, $\overline{G}_i(s) $ \cite{Nieves:2001wt}. The divergent contribution, $G_i(s_{i+})$, can be regularized, either by
one subtraction at certain scale ($\sqrt{s}=\mu$) %
\begin{eqnarray}
G_i^\mu(s) =\overline{G}_i(s) - \overline{G}_i(\mu^2) ,
\label{eq:relation}
\end{eqnarray}
or  using a sharp-cutoff regulator $\Lambda$ in momentum space, so  that
\begin{equation}
G^{\Lambda}_i(s) =\overline{G}_i(s) + G_i^{\Lambda}(s_{i+}). \label{eq:uvcut2}
\end{equation}
 If one uses channel-dependent cutoffs, the one-subtraction RS  is recovered by
choosing  $\Lambda_i$ in each channel in such a way that
\begin{equation}
G^{\Lambda_i}_i(s_{i+})= -\overline{G}_i(\mu^2) .
\label{eq:subtraction}
\end{equation}
Nevertheless, if one uses a common UV cutoff in a given $CSI$ sector, both RSs are independent  leading to
different results.

The different dynamically-generated excited $\Omega_c^{(*)}$ are obtained as poles of the
scattering amplitudes in each $J$ sector for $C=1,S=-2, I= 0$ (see Refs.~\cite{Romanets:2012hm,Nieves:2017jjx} for details on the $\Omega_c^{(*)}$ sector). 


\section{Excited $\Omega_c^{(*)}$ states}
\label{excite}

Five new narrow excited $\Omega_c^0$ states were
identified by the LHCb Collaboration in the $\Xi_c^+ K^-$ spectrum in $pp$
collisions: the $\Omega_c^0(3000)$, $\Omega_c^0(3050)$,
$\Omega_c^0(3066)$, $\Omega_c^0(3090)$ and the $\Omega_c^0(3119)$, the
last three also observed in the $\Xi_c^{'+} K^-$ decay. Also, a 
broad resonance with mass of 3188 MeV was seen in the $\Xi_c^+ K^-$ spectrum. 

\subsection{One-subtraction regularization}

\begin{table}[t]
  \centering
  \caption{ $\Omega_c^{(*)}$ excited states as found in
    Ref.~\cite{Romanets:2012hm}. We label them from {\bf a} to {\bf e}, according to their energy position. This table is taken from \cite{Nieves:2017jjx}.}
  \label{tab:table1}
  \begin{tabular}{c|c|c|c|}
      Name & $M_R$ (MeV) & $\Gamma_R$ (MeV) & $J$  \\
    \hline
    {\bf a} & 2810.9 & 0 & 1/2  \\
    \hline
    {\bf b} & 2814.3 & 0 & 3/2  \\
    \hline
    {\bf c} & 2884.5 & 0 & 1/2  \\
    \hline
    {\bf d} & 2941.6 & 0 & 1/2  \\
    \hline
   {\bf e} & 2980.0 & 0 & 3/2  \\
  \end{tabular}
\end{table}

In Ref.~\cite{Romanets:2012hm} five excited
 $\Omega_c$ states with spin-parity $J=1/2^-$ and $J=3/2^-$  were found, with masses
below 3 GeV (Table~\ref{tab:table1}). Given the mass, it is difficult to identify them with the LHCb results.

In order to obtain these five excited
 $\Omega_c$ states, the baryon-meson loops were regularized with one-substraction at the scale $\mu = \sqrt{\alpha \left(m_{th}^2+M_{th}^2 \right)}$, with $\alpha=1$, whereas $m_{th}$ and $M_{th}$ are the masses of the meson and baryon of
the channel with the lowest threshold in the given $CSI$ sector~\cite{Hofmann:2005sw}. It is then possible to change slightly the subtraction point by changing  $\alpha$. We obtain that for $\alpha=1.16$ the states {\bf d} and {\bf e}  in Table~\ref{tab:table1} can be now located near the experimental
$\Omega_c(3000)$ and $\Omega_c(3050)$ (see Table \ref{tab:table2}).  Whereas the state with mass 2999.9 MeV is mainly generated by  $\Xi_c^{'+} \bar K$ , the  3036.3 MeV state has a dominant $\Xi_c^* \bar K$  component. By allowing $\Xi_c^* \bar K \to
\Xi_c \bar K$ $d-$wave transition we can reconcile our results with the experimental decay $\Xi_c^+ K^-$.

\begin{table}[t]
  \centering
  \caption{  $\Omega_c^{(*)}$ excited states using one-subtraction regularization with $\alpha=1.16$, taken from \cite{Nieves:2017jjx}.}
  \label{tab:table2}
  \begin{tabular}{c|c|c|c||c|c}
      Name & $M_R$ (MeV) & $\Gamma_R$ (MeV) & $J$ & $M_R^{exp}$ & $\Gamma_R^{exp}$ \\
    \hline
    {\bf a} & 2922.2 & 0 & 1/2 & --- & --- \\
    \hline
    {\bf b} & 2928.1 & 0 & 3/2 & --- & --- \\
    \hline
   {\bf c} & 2941.3 & 0 & 1/2 & --- & --- \\
    \hline
    {\bf d}& 2999.9 & 0.06 & 1/2 & 3000.4 & 4.5 \\
    \hline
    {\bf e} & 3036.3 & 0 & 3/2 &  3050.2 & 0.8 \\
  \end{tabular}
\end{table}

In order to study the dependence of our results in the regularization scheme in a controlled manner, we study a different RS. Therefore, we employ a common UV
cutoff for all baryon-meson loops within a reasonable range. In this way, we avoid any uncontrolled reduction of any
baryon-meson channel, while preventing an arbitrary change of the subtraction constants.

\subsection{Common UV cutoff regularization}

\begin{table}[t]
  \centering
  \caption{ $\Omega_c^{(*)}$ excited states using a common UV cutoff regularization  of $\Lambda=1090$
    MeV,  taken from \cite{Nieves:2017jjx}. }
  \label{tab:table3}
  \begin{tabular}{c|c|c|c||c|c}
      Name & $M_R$ (MeV) & $\Gamma_R$ (MeV) & $J$ & $M_R^{exp}$ & $\Gamma_R^{exp}$ \\
    \hline
    {\bf a} & 2963.95 & 0.0 & 1/2 & --- & --- \\
    \hline
    {\bf c} & 2994.26 & 1.85 & 1/2 & 3000.4 & 4.5 \\
    \hline
    {\bf b} & 3048.7 & 0.0 & 3/2 & 3050.2  &  0.8 \\
    \hline
    {\bf d} & 3116.81 & 3.72 & 1/2 & 3119.1/ 3090.2  & 1.1/ 8.7   \\
    \hline
    {\bf e} & 3155.37 & 0.17 & 3/2 &  --- & --- \\
  \end{tabular}
\end{table}

First we have to be able to follow the original $\Omega_c^{(*)}$ in the
complex energy plane as we modify our prescription from one-subtraction to a common UV cutoff regularization for the computation
of the subtraction constants. Thus, we vary the loop function for each channel by
\begin{equation}
G_i(s) = \overline{G}_i(s)-(1-x) \overline{G}_i(\mu^2)+x G_i^{\Lambda}(s_{i+}),
\end{equation}
with $x$ being a parameter that changes adiabatically from $0$ to $1$, and
$\mu^2=(m_{th}^2+M_{th}^2)$.

Our results for $\Omega_c^{(*)}$  for a cutoff of $\Lambda=1090$ MeV are shown in Table~\ref{tab:table3}. Three poles (named {\bf c},
{\bf b} and {\bf d}) can be identified with the three experimental
states at 3000 MeV, 3050 MeV, and 3119 or 3090 MeV, respectively. This is because of the closeness in energy to the experimental
states and  the dominant contribution from the experimental $\Xi_c \bar K$ and $\Xi_c^{'} \bar K$ channels in the dynamical generation of the states.
Moreover, the states  {\bf b} with $J=3/2$ and
{\bf c} with $J=1/2$ in Table~\ref{tab:table3} for $\Lambda=1090$ MeV would belong to the same  ${\rm SU(6)}_{\rm lsf}$ $\times$ HQSS multiplets as the  $\Lambda_c(2595)$ and
$\Lambda_c(2625)$, or the $\Lambda_b(5912)$ and $\Lambda_b(5920)$.

\begin{figure*}[tbh]
  \includegraphics[width=\columnwidth,height=\columnwidth]{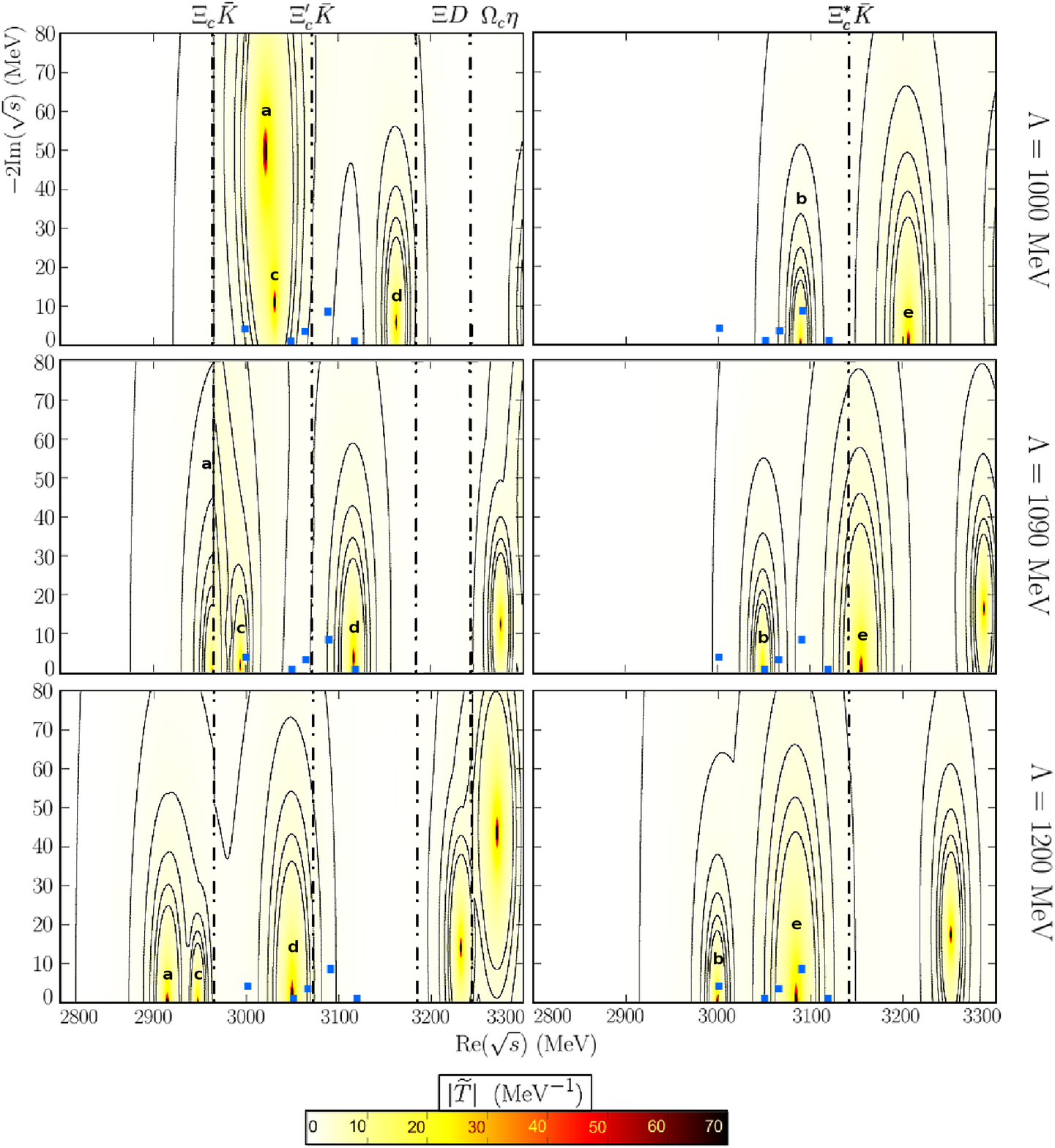}
  \caption{$\Omega_c^{(*)}$ excited states for different 
    cutoffs.  The blue squares show the experimental
    results, while dashed-dotted lines indicate the closest baryon-meson
    thresholds. The left panels are for $J=\frac{1}{2}$ and the right
    ones for $J=\frac{3}{2}$. For the two largest values of
    $\Lambda$, some states from less attractive ${\rm
      SU(6)}_{\rm lsf}\times$HQSS multiplets are also visible for higher masses. This figure is taken from \cite{Nieves:2017jjx}.}
  \label{fig:lambdas}
\end{figure*}

Now, we have to determine the dependence of our dynamical generated states on the value of UV cutoff. Thus, higher and lower values than $\Lambda=1090$ MeV are considered, approximately 100 MeV apart from $\Lambda= 1090$ MeV, as seen in Fig.~\ref{fig:lambdas}. Much higher or lower values of the cutoff will not generate states in the experimental mass region.  From Fig.~\ref{fig:lambdas}  we can conclude  that  (probably) at least three of the experimental states can be identified with three of our $\Omega_c^{(*)}$. 

As mentioned in the Introduction, the molecular nature of the five experimental $\Omega_c$ has been also revisited within other molecular schemes. In Ref.~\cite{Montana:2017kjw} two $J=1/2$ molecular states were
identified with the experimental $\Omega_c(3050)$ and
$\Omega_c(3090)$. These two $J=1/2$ $\Omega_c$ states were
reproduced in Ref.~\cite{Debastiani:2017ewu}, due to the use of the same interaction in the $J=1/2$ sector. However, 
in Ref.~\cite{Debastiani:2017ewu}  a $J=3/2$ molecular state was also identified 
as the experimental $\Omega_c(3119)$, since the model allows for the interaction of baryon $3/2^+$ and pseudoscalar mesons.
Compared to these works, our model for $\Lambda=1090$ generates $J=1/2^-$ $\Omega_c(3000)$, $\Omega_c(3119/3090)$ and $J=3/2^-$ $\Omega_c(3050)$. The difference between the predictions of our model and those previous schemes lies in the use of a different RS as well as  different  baryon-meson interaction matrices, in particular for the channels involving $D$, $D^*$ and light vector mesons.

The broad structure around 3188 MeV determined by LHCb has also been studied in Ref.~\cite{Wang:2017smo}. The authors have indicated that it could be interpreted as the superposition of two $D \Xi$ bound states. In our case, we cannot make any identification, since most probably this wide state would come from a less attractive ${\rm SU(6)}_{\rm lsf}\times$HQSS representation. Furthermore, a loosely bound molecule of mass  3140 MeV was predicted in Ref.~ \cite{Chen:2017xat}. We cannot associate any of our states to this one, since the authors in \cite{Chen:2017xat}  did not consider $\Xi^{(*)}D^{(*)}$ channels.

\section{Conclusions}

We have revisited our previous work on the $\Omega_c^{(*)}$ states of Ref.~\cite{Romanets:2012hm} in view of the recent experimental results by the LHCb (and Belle) Collaboration. In this previous paper, five odd-parity $\Omega_c^{(*)}$ states were dynamically generated with masses below 3 GeV. By  implementing a different RS, we find that the some of the predicted masses can be moved up in energy closer to the experimental energy region.  Indeed, we implement two different RS schemes and analyze the consequences for the mass, width and dominant decay channels of the dynamically generated $\Omega_c^{(*)}$ states. We  conclude that  (probably) at least three of the states observed by LHCb have spin-parity $J=1/2^-$ and $J=3/2^-$.

\section*{Acknowledgments}
L.T. acknowledges support from the Heisenberg Programme (DFG) - Project Nr. 383452331, the THOR COST Action CA15213 and the DFG through the grant CRC-TR 211. R. P. Pavao wishes to thank
the Generalitat Valenciana in the program Santiago Grisolia. This
research is supported by the Spanish Ministerio de Ciencia, Innovaci\'on y Universidades and the European Regional Development Fund, under
contracts FIS2014-51948-C2-1-P, FIS2017-84038-C2-1-P, FPA2016-81114-P and SEV-2014-0398, and by Generalitat Valenciana under
contract PROMETEOII/2014/0068.


\begin{thebibliography}{9}

\bibitem{Aaij:2017nav}
R. Aaij et al. [LHCb Collaboration],  Phys. Rev. Lett. \textbf{118}, 182001 (2017)  

\bibitem{Yelton:2017qxg}
J. Yelton et al. [Belle Collaboration], Phys. Rev. D \textbf{97}, 051102 (2018)  
 
 \bibitem{JimenezTejero:2009vq}
 C. E. Jimenez-Tejero, A. Ramos and I. Vidana,  Phys. Rev. C \textbf{80},  055206 (2009) 
 
  \bibitem{Hofmann:2005sw}
  J. Hofmann and M. F. M. Lutz,  Nucl. Phys. A \textbf{763}, 90 (2005) 
  
  \bibitem{GarciaRecio:2008dp}
   C. Garcia-Recio, V. K. Magas, T. Mizutani, J. Nieves, A. Ramos, L. L. Salcedo et al., Phys. Rev.
D \textbf{79}, 054004 (2009)  
 
 \bibitem{Romanets:2012hm}  
 O. Romanets, L. Tolos, C. Garcia-Recio, J. Nieves, L. L. Salcedo and R. G. E. Timmermans,
 Phys. Rev. D \textbf{85}, 114032 (2012)
   
   \bibitem{Montana:2017kjw}
   G. Montana, A. Feijoo and A. Ramos, Eur. Phys. J. A \textbf{54}, 64 (2018) 
   
   \bibitem{Debastiani:2017ewu}
 V. R. Debastiani, J. M. Dias, W. H. Liang and E. Oset, Phys. Rev. D \textbf{97}, 094035 (2018)  
   
   \bibitem{Gamermann:2010zz}
 D. Gamermann, C. Garcia-Recio, J. Nieves, L. L. Salcedo and L. Tolos, Phys. Rev. D \textbf{81}, 094016 (2010)  
   
   \bibitem{GarciaRecio:2012db}
C. Garcia-Recio, J. Nieves, O. Romanets, L. L. Salcedo and L. Tolos, Phys. Rev. D \textbf{87}, 034032 (2013) 
   
   \bibitem{Garcia-Recio:2013gaa}
C. Garcia-Recio, J. Nieves, O. Romanets, L. L. Salcedo and L. Tolos,  Phys. Rev. D \textbf{87}, 074034 (2013)  
   
   \bibitem{Tolos:2013gta}
 L. Tolos,  Int. J. Mod. Phys. E \textbf{22}, 1330027 (2013)  
   
   \bibitem{Garcia-Recio:2015jsa}
C. Garcia-Recio, C. Hidalgo-Duque, J. Nieves, L. L. Salcedo and L. Tolos,  Phys. Rev. D \textbf{92}, 034011 (2015)  
   
   \bibitem{Aaij:2012da}
 R. Aaij et al. [LHCb Collaboration], Phys. Rev. Lett. \textbf{109},  172003 (2012)
   
   \bibitem{Nieves:2017jjx}
J. Nieves, R. Pavao and L. Tolos,  Eur. Phys. J. C \textbf{78}, 114 (2018)  
   
   \bibitem{Nieves:2001wt}
 J. Nieves and E. Ruiz Arriola, Phys. Rev. D \textbf{64},  116008 (2001) 
   
   \bibitem{Wang:2017smo}
 C. Wang, L.-L. Liu, X.-W. Kang, X.-H. Guo and R.-W. Wang,  Eur. Phys. J. C \textbf{78},  407  (2018)
   
   \bibitem{Chen:2017xat}
R. Chen, A. Hosaka and X. Liu,  Phys. Rev. D \textbf{97}, 036016 (2018)  

\end{thebibliography}

\end{document}